\documentclass[preprint1]{emulateapj}
\usepackage{psfig}   
\usepackage{graphicx}
\usepackage{subfigure}

\newcommand{\Msun}{\ensuremath{\rm{M}_{\odot}}}

\shorttitle{Dynamical mass segregation on a very short timescale}
\shortauthors{R.~J.~Allison et~al.}

\begin{document}


\title{Dynamical mass segregation on a very short timescale}


\author{Richard~J.~Allison,\altaffilmark{1}
Simon~P.~Goodwin,\altaffilmark{1} Richard~J.~Parker,\altaffilmark{1}
Richard~de~Grijs,\altaffilmark{1,2} \\
Simon~F.~Portegies~Zwart,\altaffilmark{3} 
M.~B.~N.~Kouwenhoven\altaffilmark{1}}

\altaffiltext{1}{Department of Physics and Astronomy, University of
Sheffield, Sheffield, S3 7RH, UK; r.allison@sheffield.ac.uk}
\altaffiltext{2}{NAOC-CAS, Beijing,China} 
\altaffiltext{3}{Leiden Observatory, P.O. Box 9513, NL-2300 RA Leiden, 
The Netherlands}


\begin{abstract}
We discuss the observations and theory of star cluster formation to
argue that clusters form dynamically cool (subvirial) and with
substructure.  We then perform an ensemble of simulations of cool,
clumpy (fractal) clusters and show that they often dynamically mass
segregate on timescales far shorter than expected from simple
models. The mass segregation comes about through the production of a
short-lived, but very dense core.  This shows that in clusters like
the Orion Nebula Cluster the stars $\geq 4$\Msun can dynamically mass
segregate within the current age of the cluster. Therefore, the
observed mass segregation in apparently dynamically young clusters
need not be primordial, but could be the result of rapid and violent
early dynamical evolution.
\end{abstract}





\section{Introduction}

Most stars are observed to form in clusters (probably 70---90\%; Lada
\& Lada 2003.)\nocite{lada03}  Therefore, understanding cluster
formation is the key to understanding most star formation.
Unfortunately, observations of very young clusters ($\ll 1$~Myr) are
hampered by extinction due to those clusters still being embedded in
their natal molecular clouds.  Most often we are restricted to
observing older clusters which is problematic as these clusters may
well have undergone significant dynamical evolution.  In particular,
young clusters seem to expand significantly in the first few Myr
\citep{bastian08}, possibly due to the effects of multiple stellar
encounters \citep{vandenberk07} or by the rapid expulsion of
primordial gas \citep{goodwin06}.

A potentially significant observation is that many young clusters
appear `mass segregated,' i.e., the most massive stars are
concentrated towards the center of the cluster (Hillenbrand \&
Hartmann 1998; de Grijs et al. 2002a,b,c; Gouliermis et al. 2004).
\nocite{hillenbrand98,degrijs02a,degrijs02b,degrijs02c,gouliermis04}
Clearly, mass segregation could be primordial -- clusters may form
with the most massive stars concentrated at or near the center.
Alternatively, mass segregation could be dynamical -- the most massive
stars migrate into the center of the cluster after formation due to
two-body interactions.  As well as potentially providing constraints
on clustered star formation, the origin of mass segregation may help
distinguish models of massive star formation.  In particular, are the
masses of the most massive stars set by the mass of the core in which
they form \citep[e.g.,][]{krumholz07}, or by competitively accreting
mass due to a favorable position in the cluster (e.g., Bonnell et
al. 1998; see also Krumholz et al. 2005, and Bonnell \& Bate
2006\nocite{bonnell98b,krumholz05,bonnell06})?  If mass segregation is
primordial it might well argue in favor of competitive accretion, as
in that scenario massive stars should form in the center of a cluster.

\citet{bonnell98} investigated the evolution of virialized and
subvirial (thus collapsing) smooth Plummer spheres. They showed that
dynamical mass segregation cannot occur rapidly enough in these
situations to explain the observations of mass segregation in young
(few Myr old) clusters such as the Orion Nebula Cluster (ONC).  In
this letter we will examine the dynamical evolution of initially cool
and clumpy (fractal) stellar distributions. We will show that a
combination of subvirial velocities {\em and} substructure can often
lead to dynamical mass segregation on a timescale that is much shorter
than would usually be expected.  In $\S~2$ we argue that cool, clumpy
initial conditions are realistic -- indeed, they are both observed and
expected from theory.  In $\S~3$ we demonstrate that these initial
conditions can lead to rapid dynamical mass segregation.  In $\S~4$ we
discuss the implications of this result and draw our conclusions.

\section{The initial conditions of star clusters}
\label{sec:rev}

In this section we review the observations and theory of star cluster
formation in molecular clouds.  Our aim is to show that both
observations and theory lead us to expect that clusters form both
subvirial and with substructure.

\noindent \textit{Star clusters are initially substructured.}  Stars
form from dense cores in molecular clouds
\citep[e.g.,][]{ward-thompson07}.  Molecular clouds are observed to
have significant levels of substructure in both density and kinematics
(Williams 1999; Williams et al. 2000; Carpenter \& Hodapp
2008)\nocite{williams00,carpenter08,williams99}.  This is not
surprising, as we believe that molecular cloud structure is dominated
by supersonic turbulence which will produce complex structures (see,
Mac Low \& Klessen 2004; Ballesteros-Paredes et al. 2007, and
references therein\nocite{maclow04,ballesteros-paredes07}). It is not
surprising, therefore, that very young ($<1$ Myr) star clusters also
show significant levels of substructure (Larson 1995; Testi et
al. 2000; Elmegreen 2000; Lada \& Lada 2003; Gutermuth et al. 2005;
Allen et al. 2007)
\nocite{larson95,elmegreen00,testi00,lada03,gutermuth05,allen07}.
Detailed statistical studies also show the presence of substructure in
young clusters, and that substructure is rapidly erased
\citep{cartwright04,schmeja08}.

\noindent \textit{Star clusters form subvirial (cool).} There is
increasing evidence that star clusters are born subvirial.  Prestellar
cores are often observed to move subsonically (Belloche et al. 2001;
Andr\'e 2002; Walsh et al. 2004; Peretto et al. 2006; Kirk et
al. 2007)\nocite{belloche01,andre02,walsh04,peretto06,kirk07}.  In
hydrodynamic simulations of cluster formation starting from
dynamically hot (globally unbound) gas the stars that form also appear
to have subvirial motions (see e.g. Bate et al. 2003; Bonnell et
al. 2003).  Proszkow et al. (2009) also find that the Orion star
forming region shows signatures of subvirial dynamics on scales of
about 10~pc (see also Adams et al. 2006).  Such subvirial motions
might well be expected from a turbulent model of star formation in
clouds where stars form in converging/colliding flows (see references
above, also Adams et al. 2006).  It is also worth noting that
sub-virial initial conditions are required to erase substructure as
rapidly as is observed (Goodwin \& Whitworth 2004).

\section{The evolution of cool, fractal clusters}
\label{sec:sims}

In this section we simulate clumpy (fractal), cool clusters and show
that they mass segregate dynamically in a time similar to the crossing
time of the initial (or final) system.

\subsection{Initial Conditions}
\label{ssec:init}

We simulate the dynamical evolution of star clusters containing 1000
single stars with masses sampled from a three-part power-law mass
function \citep{kroupa02}, with minimum and maximum masses of
0.08\Msun and 50\Msun, respectively. The stars initially have a
fractal distribution within a sphere of radius 1~pc, with initial
velocities such that nearby stars have similar velocities, as
described in detail by \citet{goodwin04}. We neglect the effects of
stellar evolution because of the short duration of the simulations (4
Myr).

In this letter, we restrict our investigations to clusters with a
fractal dimension of 1.6 (giving a very clumpy distribution, where 3.0
gives a uniform sphere) and a virial ratio of $Q = 0.3$, where $Q$ is
the ratio of the kinetic to the (modulus of the) potential energy (so
that $Q=1/2$ is virialized).  Such initial conditions give the most
extreme dynamical evolution and the most rapid dynamical mass
segregation.  We will consider a much fuller range of parameter space
in a follow-up paper (R.~J.~Allison et al., in prep.).  Here, we just
wish to illustrate that rapid dynamical mass segregation can occur
with plausible initial conditions, and how this happens. The
simulations were carried out using the {\sc kira} integrator in {\sc
starlab}. \citep{portegies_zwart01}.

\subsection{Results}

\begin{figure}
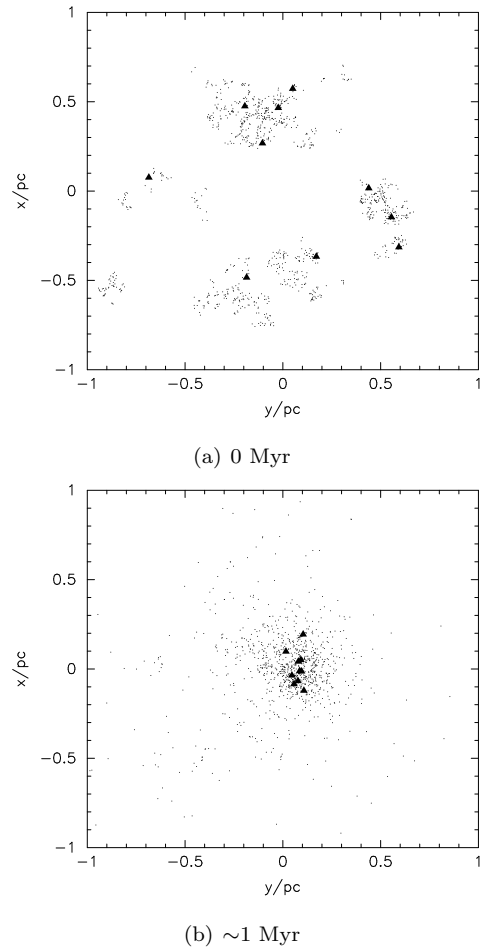

  \begin{center}
    \setlength{\subfigcapskip}{10pt}
\subfigure[0 Myr]{\label{fig:a1.21.0myr}
\includegraphics[scale=0.3,angle=270]{pos.f1000.a1.21.001.ps}}
\subfigure[$\sim$1 Myr]{\label{fig:a1.21.1myr}
\includegraphics[scale=0.3,angle=270]{pos.f1000.a1.21.083.ps}}
  \end{center}
  \caption[bf]{Typical stellar distributions at 0 and $\sim$1 Myr. The
    triangles indicate the positions of the 10 most massive stars. In
    this run the most massive and tenth most massive stars have masses
    of 23\Msun and 3.5\Msun, respectively.}
  \label{fig:a1.21}
\end{figure}

Figures~\ref{fig:a1.21.0myr} and~\ref{fig:a1.21.1myr} show the stellar
distributions initially, and after $\sim$ 1 Myr of dynamical
evolution, respectively. A comparison of the plots clearly shows that
the cluster has evolved from a clumpy and non-mass segregated state to
one which has erased substructure and appears to be mass segregated. 

We apply the method of \citet{allison09}, which compares the minimum
spanning trees (MSTs) of high-mass stars to those of a random
selection of stars to produce a quantitative measure of mass
segregation.  If the MST of the $N$ most massive stars is
significantly shorter than that of a number of sets of $N$ random
stars then the cluster is mass segregated.  The degree of mass
segregation can be quantified by the ratio of the lengths of the
average randomly selected star MST to the most massive star MST,
$\Lambda$ \citep[see][for details]{allison09}.  The greater $\Lambda$
is relative to unity, the more mass segregated a cluster is.

\begin{figure}
  \begin{center}
    \setlength{\subfigcapskip}{10pt}
    \includegraphics[scale=0.7]{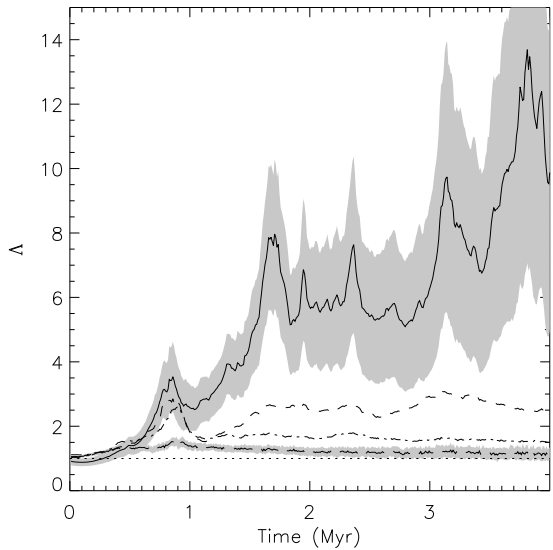}
  \end{center}
  \caption[bf]{The evolution of the mass segregation ratio $\Lambda$
    for the 10, 20, 50 and 100 most massive stars in the cluster (from
    top to bottom). The envelopes indicate 1$\sigma$ deviations, no
    error is shown for $N=20$ and 50 for clarity. The dotted
    line indicates a $\Lambda$ of unity, i.e., no mass segregation.}
  \label{fig:nmstplot}
\end{figure}

Figure~\ref{fig:nmstplot} shows the evolution of $\Lambda$ for four
subsets of the $N = 10, 20, 50$ and 100 most massive stars in the
cluster. The cluster is not mass segregated initially ($\Lambda = 1$),
but after 1 Myr the 10 most massive stars develop a significant level
of mass segregation ($\Lambda \sim 3$). The error bars in
Figure~\ref{fig:nmstplot} represent the instantaneous standard
deviation at each simulation snapshot.\footnote{The 10 most massive
stars are mass segregated at a similar level for more than a crossing
time after the initial violent relaxation phase. In a crossing time
the cluster can completely mix and every star in the cluster can
migrate to any other position in the cluster. This means that any
feature that remains constant over a crossing time is a real feature
and if the evolution of the cluster were also accounted for the
significance would be much greater than shown.}  In
Figure~\ref{fig:nmstplot} we can also see that the 20 and 50 most
massive stars also mass segregate, but by a much smaller
amount. Beyond the 50 most massive stars little mass segregation is
seen.

The cluster illustrated in Figure~\ref{fig:a1.21} is a fairly typical
example of the evolution of cool, highly fractal clusters: a collapse
from the cool initial state, which erases substructure and also
imprints mass segregation.  Unfortunately, and unavoidably, using
fractal clusters introduces a certain degree of randomness as each
fractal -- while formally the same (i.e., the same fractal dimension
and virial ratio) -- is very different in its initial distribution.
Thus, when dealing with fractals, it is vital to perform a large
ensemble of simulations.

We simulate 50 different clusters, varying only the random number seed
used to create the fractal and initial mass function.  Of these 50
simulations, 29 mass segregate within 1~Myr, and 44 show mass
segregation within 4~Myr.  Only 6 show no significant mass segregation
by the end of the simulations at 4~Myr. Here, mass segregation is
defined as any mass segregation event that lasts for longer than $\sim
0.1$ Myrs and for which $\Lambda$ has a significance $>1$.

\subsection{Mass segregation mechanism}

The mass segregation observed in these simulations is from a wholly
dynamical origin, and arises from the collapse of the cluster. The
initial conditions used in these simulations place the cluster far
from equilibrium. The cluster undergoes a gravitational collapse and
violent relaxation phase very early on in its evolution in an attempt
to virialize itself.

This collapse creates a dense core containing roughly half the 
mass in a radius of only $\sim 0.1$~pc.  The dense core only lasts for 
$0.1$ -- $0.2$~Myr, but due to its small size, this is around 
$10$ -- $20$ crossing times of the core, making it dynamically old.

Whilst $10$ -- $20$ crossing times is not enough time to reach full
equipartition, it is sufficient time to mass segregate the most
massive stars.  \citet{spitzer69} showed that the mass segregation
timescale for a star of mass $M$, $t_{\rm seg}(M)$, is

\begin{equation}
t_{\textrm{seg}}(M) \approx \frac{\langle m \rangle}{M}t_{\textrm{relax}},
\label{eq:tseg}
\end{equation}

\noindent where $\langle m \rangle$ is the average mass of a star in
the cluster ($\sim 0.4 M_\odot$ for a typical initial mass function),
and $t_{\rm relax}$ is the two-body relaxation timescale of the
cluster.  The two-body relaxation timescale depends on the number of
stars in the cluster, $N$, and the crossing time, $t_{\rm cross}$, as

\begin{equation}
t_{\textrm{relax}} \approx  \frac{N}{8\ln N}t_{\textrm{cross}},
\label{eq:trelax}
\end{equation}

\noindent where the crossing time is simply the radius of the cluster
$R$ divided by the average velocity of a star, $\sigma$.

Thus, Eq.~\ref{eq:tseg} can be rewritten as

\begin{equation}
t_{\textrm{seg}} \approx \frac{\langle m \rangle}{M}
\frac{N}{8\ln{N}}\frac{R}{\sigma}.
\label{eq:tdense.exp}
\end{equation}

Typical values for these parameters for the dense cores are $N \sim
300-500$, $R \sim 0.1$ -- $0.2$~pc, $\langle m \rangle = 0.4 M_\odot$,
and $\sigma \sim 2$~km s$^{-1}$.  The lifetimes of the dense cores
during which they can mass segregate are $t_{\rm seg} \sim 0.1$~Myr.
This suggests that clusters will be able to mass segregate above $M
\sim 2$ -- $4 M_\odot$.

In Figure~\ref{fig:nmstplot} we show that below about the 50th most
massive star there is no further mass segregation.  In this
simulation, the mass of the 50th most massive star is $\sim 2 M_\odot$
-- in good agreement with the calculation presented above.

Interestingly, \citet{allison09} find that the ONC appears to be mass
segregated down to $\sim 5 \Msun$ but not below that mass.
\citet{hillenbrand98} also find evidence for mass segregation around
5\Msun, and \citet{moeckel09} find that they cannot explain the
observations of the ONC unless only the most massive stars are mass
segregated.  This is exactly the situation predicted in our
simulations -- mass segregation down to a few solar masses, but not
below.  Indeed, observations of the mass down to which a cluster is
mass segregated (and the dynamical age of the cluster) should provide
constraints on the density and duration of the dense phase undergone
by the cluster and hence the initial conditions of the cluster.

The mass segregation seen in these simulations is due to the collapse
of the cluster and formation of the dense core. Some early, low level,
mass segregation is observed in the simulations, due to the dynamical
evolution and merger of the sub-clumps in the initial distribution
\citep{mcmillan07}, but the presence of long lived, high level mass
segregation is due almost entirely to the later evolution of the dense
core, not to prior mass segregation in clumps.

The reason that clumpy subvirial clusters are able to mass segregate
whilst smooth subvirial clusters do not (such as those simulated by
Bonnell \& Davies 1998) is due to clumpy clusters being able to
collapse to a far denser state than smooth clusters.  The potential energy, 
$\Omega$, of a cluster of mass $M_{\rm clus}$ and radius $R$ is given by
\[
\Omega = \alpha \frac{G M_{\rm clus}^2}{R}
\]
where $\alpha$ is a structure parameter (for example in  Plummer sphere 
if $R$ is the Plummer radius then $\alpha \sim 0.75$).

For a cluster with an initial potential energy $\Omega_0$ 
(with radius $R_0$ and structure parameter $\alpha_0$),
and a final potential energy $\Omega_f$ (with radius $R_f$ and 
structure parameter $\alpha_f$) the initial and
final potential energies are related to the initial virial ratio $Q_0$ by
\[
\Omega_0 (1 - Q_0) = \frac{\Omega_f}{2},
\]
assuming the cluster ends virialized. The ratio of initial-to-final radii is
\[
\frac{R_0}{R_f} = \frac{\alpha_0}{\alpha_f} 2 (1 - Q_0).
\]
In the case where a cluster starts smooth (e.g. a Plummer sphere) it
will retain this structure and so $\alpha_0 \sim \alpha_f$ and so if a
cluster starts with a virial ratio of $Q_0 = 0.3$ it will only
collapse by a factor of $\sim 1.4$.  This is not enough to
significantly increase the speed of mass segregation.  This is
consistent with the lack of mass segregation found by Bonnell \&
Davies (1998) in the collapse of a $Q_0 = 0.1$ Plummer sphere (which
would collapse by a factor of $\sim 1.8$).

However, when a clumpy cluster collapses it erases substructure and
becomes smooth.  A reasonable estimate for the final structure
parameter is that of a Plummer sphere with $\alpha_f \sim 0.75$.  But
the highly fractal initial conditions have a structure parameter which
is significantly larger than that for a smooth distribution. For a
fractal with fractal dimension = 1.6 numerical experiments show that
$\alpha_0 \approx 1.5$ (this is an approximate figure and can change
dramatically depending on the random number seed of the fractal).
This means that the degree of collapse to virialise from a realistic
initial virial ratio of $Q_0=0.3$ is a factor of about 2.5, from 1pc
initially to around 0.4 pc, which is consistent with our simulations
(note that the core radius will be significantly smaller than 0.4~pc).

It is important to note that clusters which undergo a collapse and
dynamically mass segregate will often be significantly {\em
dynamically} older than their current crossing time might suggest
\citep[as also emphasized by][]{bastian08}. Clusters can dynamically
segregate in $1$ -- $2$~Myr even if this is comparable to their
initial or current crossing times (as inferred from their sizes)
because they have undergone a dense phase, which means that they are
actually many crossing times old (in their cores at least).

The dense cores are very short-lived as the low-mass stars expand due
to the transfer of energy from the high-mass stars to low-mass stars
during the mass segregation.  The lifetimes of the dense phase is
usually only 0.1 -- 0.2~Myr.

There is one obvious omission from our simulations -- gas.  For
simplicity, we neglect the background potential of the gas from which
the star clusters formed.  However, we believe that this omission will
not affect our results significantly.  Much of the gas that does not
form stars has a high velocity dispersion: the cloud as a whole may
well be unbound due to supersonic turbulence.  Stars are able to form
cool as they form in converging and shocked regions within a globally
unbound cloud (e.g. Bate et al. 2003; Bonnell et al. 2003).
Therefore, we would not expect significant amounts of gas to collapse
with the stars (obviously some will, but the contribution to the
potential will become small as the stars collapse to a denser
configuration).  This is observed in the simulations of \citet{bate09},
where gas which has not formed stars has migrated away from the star
forming areas due to its initial velocity dispersion.  In
particular, Fellhauer, Wilkinson, \& Kroupa (2009) show that the
presence of a background potential does not significantly alter the
merging behavior of subclumps.

\section{Conclusions}

Observations and theory suggest that many young star clusters form
with a significant amount of substructure.  Observations also show
that young clusters lose their substructure on a timescale comparable
to their current crossing times ($1$ -- $2$~Myr).  Simulations suggest that
the only way in which this could happen is if clusters are born
dynamically cool \citep{goodwin04}.  Therefore we
argue that the correct initial conditions of many star clusters must
be a cool, clumpy distribution.

We conduct an ensemble of simulations of cool, clumpy clusters (with
fractal dimension $1.6$ and virial ratio $0.3$) to investigate the
early dynamical evolution of such clusters.  We also find that cool,
fractal clusters tend to dynamically mass segregate down to a few
solar masses during a dense, but short-lived, state at the end of
their collapse.  Such limited mass segregation of only the most
massive stars is what appears to be observed in the ONC
\citep{allison09,moeckel09}.

Such rapid dynamical mass segregation in clusters with realistic
initial conditions shows that the most massive stars do not have to
form in the centers of clusters for very young clusters to be mass
segregated.  This shows that massive stars could form in relative
isolation in large cores and mass segregate later, possibly avoiding
the need for competitive accretion to form the most massive stars in
the center of a cluster \citep{bonnell01}.  However, competitive
accretion may still play an important role, further increasing the
masses of the most massive stars if they mass segregate while gas is
still present.

\acknowledgements

RJA and RJP acknowledge financial support from STFC. SPZ is grateful
for the support of the Netherlands Advanced School in Astrophysics
(NOVA), the LKBF, and the Netherlands Organization for Scientific
Research (NWO). MBNK was supported by PPARC/STFC under grant number
PP/D002036/1. This work has made use of the Iceberg computing
facility, part of the White Rose Grid computing facilities at the
University of Sheffield. We acknowledge the support and hospitality of
the International Space Science Institute in Bern, Switzerland, where
part of this work was done as part of a International Team Programme.

\end{document}